\documentclass[fleqn,usenatbib]{mnras}

\usepackage[T1]{fontenc}
\usepackage{url}
\DeclareRobustCommand{\VAN}[3]{#2}
\let\VANthebibliography\thebibliography
\def\thebibliography{\DeclareRobustCommand{\VAN}[3]{##3}\VANthebibliography}

\usepackage{graphicx} 
\usepackage{amsmath} 
\usepackage{amssymb} 
\usepackage{caption}
\usepackage{subcaption}
\usepackage{longtable}
\usepackage{multirow}
\usepackage{multicol}
\usepackage{newtxtext,newtxmath}

\title[PSFs for mapping of artificial night sky luminance]{PSFs for mapping artificial night sky luminance over large territories}

\author[A. Simoneau et al.]{
Alexandre Simoneau$^{1}$,
Martin Aubé$^{1,2,3}$\thanks{E-mail: martin.aube@cegepsherbrooke.qc.ca},
Jérôme Leblanc$^{2}$,
Rémi Boucher$^{4}$,
\newauthor
~Johanne Roby$^{2}$ and Florence Lacharité$^{2}$
\\
$^{1}$Département de géomatique appliquée, Université de Sherbrooke, 2500 Boul. de l'Université, Sherbrooke, J1K 2R1, Canada\\
$^{2}$Cégep de Sherbrooke, 475 rue du Cégep, Sherbrooke, J1E 4K1, Canada\\
$^{3}$Physics Department, Bishop's University, 2600 College St.
Sherbrooke, J1M 1Z7, Canada\\
$^{4}$Parc national du Mont-Mégantic, 189 route du parc, Notre-Dame-des-Bois, J0B 2E0, Canada
}

\date{Accepted 2021 March 04. Received 2021 March 03; in original form 2020 December 09}

\pubyear{2020}

\begin{document}
\label{firstpage}
\pagerange{\pageref{firstpage}--\pageref{lastpage}}
\maketitle

\begin{abstract}
Knowledge of the night sky radiance over a large territory may be valuable information to identify sites appropriate to astronomical observations or for assessing the impacts of artificial light at night on ecosystems. Measuring the sky radiance can be a complex endeavour depending on the desired temporal and spatial resolution. Similarly, modelling of artificial night sky radiance for multiple points of a territory can represent a significant amount of computing time depending on the complexity of the model used. The use of the convolution of a point spread function with the light sources geographical distribution has been suggested in order to model the sky radiance over large territories of hundreds of kilometres in size. We determine how the point spread function is sensitive to the main driving parameters of the artificial night sky radiance such as the wavelength, the ground reflectance, the obstacles properties, the Upward Light Output Ratio and the Aerosol Optical Depth using the Illumina v2 model. The obtained functions were used to model the artificial night sky brightness of the Mont-Mégantic International Dark Sky Reserve for winter and summer conditions. The results were compared to the New world atlas of artificial night sky brightness, the Illumina v2 model and in situ Sky Quality Camera measurements. We found that the New world atlas overestimates the artificial sky brightness by 55\% whereas the Illumina model underestimates it by 48\%. This may be due to varying atmospherical conditions and the fact that the model only accounts for public light sources.
\end{abstract}




\section{Introduction}
Artificial night sky radiance (ANSR) results from light being emitted or reflected towards the sky and some of it being redirected towards the ground. Many factors such as the  aerosols or molecules scattering \citep{sciezor2020} affect the level of ANSR. Other mechanisms characterizing the environment also determine the behaviour of the ANSR. These include the atmospheric attenuation, the second order of scattering (especially important for observers far from the sources), the obstacles blocking the propagation of light in certain directions (mainly buildings and trees) and the ground reflectance redirecting light oriented towards the ground to the sky \citep{aubephil2015}. For many of these mechanisms, the spectral power distribution and angular emission of lamps are key factors.

Although ANSR has been known to impact astronomical observations for several decades \citep{burns1910, riegel1973, walker1973, falchi2016}, we now know that it can also affect wildlife and flora in various ways \citep{Plants,Bats,Fish,dungbeetles,Songbirds,turtles,wallabies,Nspecies}. 

Mapping the levels of the ANSR over large territories could be helpful in devising ways to mitigate the various negative impacts of exposure to nocturnal light. One such potential application is identifying nocturnal wildlife corridors that favour wildlife migrations and movements \citep{zeale2018}. Another possibility is to generate sky radiance maps corresponding to a lighting technology change to predict the possible impacts of a lighting conversion \citep{aube2014sky, HAWAII, aubeIllumv2}. One can also study how the light emitted from a municipality affects its surroundings \citep{bara2018, hector2020, aubeIllumv2}. As for now, some methods were developed to map the ANSR such as the New World Atlas (NWA) of Artificial Night Sky Brightness \citep{falchi2016} which uses a combination of a light propagation model with high-resolution satellite data from the Visible Infrared Imaging Radiometer Suite Day/Night Band (VIIRS-DNB) sensor \citep{VIIRS2016} on the Suomi National Polar-orbiting Partnership (NPP) satellite. The NWA covers the entire globe. Another example is a high-resolution map of light pollution based on non-physical distance relationship models (e.g., \citealt{NETZEL2018300}). But another methodology is to produce an ANSR map from satellite imagery combined with the convolution of a single source {PSF. This can then be optimized} using the Fast Fourier Transform (FFT, \citealt{fft}). \citet{Bar__2020} suggested such a methodology with an ad hoc point spread function (PSF) for a single source. Subsequently, \citet{hector2020} used the same idea with a PSF determined using the Illumina v1 model \citep{aube2005,aube2007light,aubephil2015}.

The use of satellite imagery alone for determining upward radiation has some shortcomings. Currently operational satellites for nighttime observation of earth in the visible range are limited in spatial and spectral resolution as well as in angular emission information. The spatial and spectral resolution problems can be addressed by using other sources of information such as images taken from the international space station \citep{ISS}. However, one of the main problems is the limited angular emission information which restricts the imaging to near zenith radiation, when it is known that near-horizon emissions can be a significant cause of skyglow \citep{aubephil2015}. This limitation can be resolved by using a detailed inventory of the emission properties of the light sources on the ground. Such an inventory would contain the spectral and angular emission functions of each light source on the studied territory. This study is based on that kind of public lighting infrastructure inventory.

As for this paper, we are using the latest version of the Illumina model (Illumina v2) that is considering many variables and physical processes such as ground reflectance, a wider range of wavelength and atmospheric conditions and the local angular and spectral properties of the light sources and blocking obstacles. One output of the model that is useful for this application is the contribution maps, which contains information about how much each pixel of the domain contributes to the total artificial sky brightness at a given observation site. We are using these contribution maps generated by the model to produce high-resolution PSFs while reducing the necessary computation time to its minimum. We produce theses PSFs for the parameter combinations we want to investigate and for each type of light source. It is then possible to obtain the sky radiance map by convolution of the appropriate PSFs with a map of the distribution of the light sources on the territory. The details of the process are explained in Section \ref{sec:methods}. These maps can then be integrated using any spectral function to simulate a measurement apparatus. As a demonstration case, we use this approach to make a zenith luminance map over the Mont-Mégantic International Dark Sky Reserve (IDSR) in Québec, Canada. The Mont-Mégantic IDSR is a territory of 5,275 $km^2$ \citep{ricemm}. It has an average radius of 50 km with the Mont-Mégantic observatory in its centre. It contains 34 municipalities, most of them being small villages, and the city of Sherbrooke ($\approx$ 170,000 inhabitants). The maps are compared to in-situ measurements to validate the accuracy of this approach.

\section{Methods}
\label{sec:methods}

We use the method proposed by \citet{bara2018} to compute the ANSR over a large area. The basic idea of this approach is to determine the PSF of the light sources and use it in order to evaluate the total effect of multiple light sources distributed over a large territory. In the case where the PSF is invariant by translation, this can be achieved by using a convolution of a light emission map with the PSF of the light sources. This convolution method is common in image processing applications. Moreover, convolutions can be performed efficiently by using the convolution theorem, which provides equivalent operations in Fourier space. The method is detailed in subsection \ref{sec:maths}. At the end of the process, we obtain a map of the zenith ANSR as seen from any point of a territory. In our case we converted the multispectral zenith ANSR map into zenith sky luminance map using the photopic spectral weighting function.

To obtain an accurate estimate of the sky luminance using this approach, it is important to use a representative PSF for the evaluated region. We used the Illumina v2 model \citep{Aube2018,aubeIllumv2} to determine the PSF according to a variety of combinations of environmental parameters. Subsection \ref{sec:model} describes how Illumina v2 was used for this purpose. Subsection \ref{sec:SQC} describes how in situ measurements were made with an all-sky calibrated camera.

The PSF varies depending on environmental parameters like the ground reflectance, the blocking obstacle height and transparency, the Aerosol Optical Depth (AOD), the wavelength of the light and the Upward Light Output Ratio (ULOR) \citep{aubephil2015} of the light sources. Variations of the PSF relative to most of these parameters are discussed in subsection \ref{sec:psf}. 

Finally, subsection \ref{sec:RICEMM} details how the method is applied to generate night sky luminance maps over the Mont-Mégantic IDSR and subsection \ref{sec:sqc} compares the obtained results with the in situ measurements. 

\subsection{Building artificial sky radiance maps with a PSF}
\label{sec:maths}

The method proposed by \citet{bara2018} to compute the ANSR over a large territory consists of determining the PSF for the sources on the territory and use that to evaluate the artificial sky brightness for the whole domain. This can be done in the specific case where the PSF is invariant by translation by performing two-dimensional convolutions instead of light propagation integrals as is standard with radiative transfer models such as Illumina v2. Moreover, it is possible to work in the Fourier space since a convolution integral is equivalent to a multiplication in Fourier space.

The point of using the convolution theorem is that an efficient algorithm called FFT \citep{fft} exists to transform between spatial (real) and Fourier space. The convolution using the FFT can be faster over larger domains while being equivalent to the two-dimensional convolution \citep{convolution}. Because this method of doing convolution is common, the numerical library we use has it already implemented. We therefore simply need to use the appropriate function (scipy.signal.fftconvolve).

In order to use this method, we need to make the same assumptions that \cite{Bar__2020} did. First, we assume that the properties of the atmosphere (and environment) are only varying with respect to the altitude but remain constant when it comes to horizontal translations. This implies that the topography of the domain is assumed to be flat and the obstacles, the lamps angular emission function and the ground reflectance are assumed to have uniform properties all over the territory. These assumptions make the PSF spatially invariant by translation. In other words, only the relative positions between the source and the observer matters.

The PSF depends on the wavelength as multiple physical processes involved in light propagation are wavelength dependent. Moreover, the light emissions on the territory can be decomposed in a linear combination of standard lamps. The method can then be applied independently on each of these lamp types, as the total light detected is simply the sum of all these sources.
We therefore factorize the light sources into an ULOR and wavelength dependent term, each associated with a different PSF for any set of wavelengths, ULORs and viewing angles \citep{Bar__2020}.

\subsection{Model used} 
\label{sec:model}
\begin{figure}
    \centering
    \includegraphics[width=.82\columnwidth]{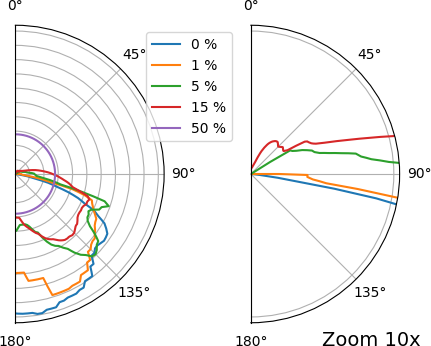}
    \caption{Angular emission functions used for the corresponding ULOR. 0° represents zenith. The right plot is a zoomed view of the emission functions. It highlights that for the 1\% ULOR function, upward light is emitted very close to the horizon.}
    \label{fig:ulor}
\end{figure}

In order to generate the PSFs required to make the ANSR map of the territory, we are using the Illumina v2 radiative transfer light pollution model \citep{aubeIllumv2}. We are simulating a set of spectral bins covering most of the visible spectrum, as well as a set of various ULORs, AODs and obstacles heights for both summer and winter typical ground reflectances. In our case, ULOR corresponds to different types of actual lamp fixtures. The exact functions used are shown in Figure~\ref{fig:ulor}. It is interesting to note that the 1\% ULOR mostly emits upward light towards the horizon.

We simulated a large flat territory ($\sim$145 $\times$ 145 km) full of white lights (constant unitary radiant flux in each spectral bin) with one observer in its centre. We used white light for simplicity as we need light emission for every wavelength to properly evaluate the PSF.


We are interested in the light perceived by the observer for each light source of the territory. This information is given by contribution maps that are routinely generated by Illumina v2. 
A similar application of the Illumina model to find PSFs was done by \citet{hector2020} using the Illumina v1 model. They differ from our usage of the model by not using the contribution maps. Instead, they performed a simulation of an observer at different distances from a light source, hence one simulation per distance. In our case, the PSFs are assumed to be spatially invariant so we get an observer per pixel of the contribution maps. Therefore, we have many observers per simulation. This is why our approach with the contribution maps is more efficient and precise to define the PSFs.

\begin{figure}
 \includegraphics[width=\columnwidth]{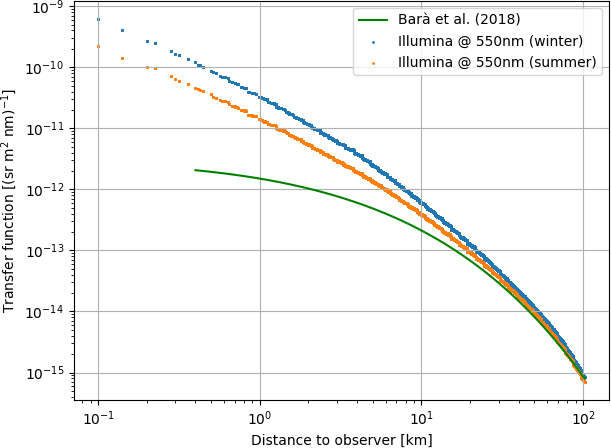}
    \caption{Summer and winter PSFs obtained with Illumina v2 for 550 nm with the reference parameters presented in Table~\ref{tab:base_case}. The green curve is the PSF used by \citep{bara2018} in their computations with an arbitrary normalization factor $c = 1.5\times10^{-12}$ to fit at the last value of our simulated domain.}
    \label{fig:psf_2d_figure}
\end{figure}

Because Illumina is a tool built for simulating ANSR and not producing PSFs, the contribution maps obtained are in units of W/m$^2$/sr/nm. We therefore need to divide the contribution maps by the radiant spectral power (which is in units of W/nm) used for modelling purposes in order to obtain "true" PSFs. Knowing the distance between a source and an observer and the radiance seen by the observer, we get a relation between radiance observed and the distance, which corresponds to a PSF (see  Figure \ref{fig:psf_2d_figure}).
In Figure \ref{fig:psf_2d_figure}, one can compare the summer and winter PSF for the reference parameters (see in Table \ref{tab:base_case}) with the one used by \citet{bara2018}. One first clear difference is that the functions that we obtained decrease more rapidly with distance. If we normalize the \citet{bara2018} at 100 km, we see that at a distance of 1 km from the source, our PSFs are more than one order of magnitude higher. That means that the Illumina v2 determined PSFs will increase the relative effect at short distances compared to the one used by \citet{bara2018}.

Since Illumina v2 outputs results by layers of different pixel resolution \citep{aubeIllumv2}, the density of values along the distance from the observer is not constant. We linearly interpolate between each point of Figure~\ref{fig:psf_2d_figure} to get a value for any source-observer distance.
Once a PSF is evaluated, we perform the convolution over the domain of interest, which is the domain of the map we wish to model the ANSR.


\subsection{SQC measurements}
\label{sec:SQC}

In situ measurements of the night sky luminance were taken using a digital single-lens reflex camera (Canon EOS 5D Mark II) and a fisheye lens (Sigma 8mm f/3.5 EX DG Circular Fisheye). All-sky images were made with the camera mounted on a tripod and pointing directly at zenith. Aperture was constant at f/3.5, with an ISO of 1600 or 3200. Exposures ranged between 15 and 180 seconds, depending on the brightness of the site and in order to maximize the signal-to-noise ratio while avoiding saturation of the sensor. When light sources were nearby, means have been taken to block any direct light from hitting the lens. All images were taken under clear sky conditions, after astronomical twilight and with the Moon below the horizon. Also, the presence of the Milky Way plane near zenith was avoided as much as practically possible, especially in the dark sites. 

The captured images were analyzed with the commercial "Sky Quality Camera"  (SQC) software, version 1.9.3, developed by Andrej Mohar from Euromix Ltd, Slovenia. Luminance data in mcd/m$^2$ is obtained for every pixel by processing of the green channel of the RAW all-sky images by the SQC software. The calibration of this specific camera and lens combination was made by the software manufacturer using standard astronomical photometry at the time of purchase. The exact spectral sensitivity curve is not available, but is stated by the manufacturer to fit very close to the CIE photopic luminosity function $V(\lambda)$. Luminance values for zenith were obtained from the software by measuring the mean value of a circular area with a radius between 0° to 10° of zenith angle. We did not apply the Milky Way and bright stars subtraction tools that are available in the software in order to include all natural light sources.
\vspace{2pt}
\section{Results}

In the next subsections, we first present PSFs with varying environmental and lamps parameters and discuss the potential dependency between these functions and said parameters (subsection~\ref{sec:psf}). Then, in subsection~\ref{sec:RICEMM}, we present the application of PSF with appropriate parameters to model sky radiance maps of the Mont-Mégantic IDSR with a given spectral sensitivity weighting function. In our specific case, the sensitivity is the photopic curve since we computed the zenith sky luminance map from the multispectral ANSR maps.

\subsection{PSFs dependency on environment and lamps parameters}
\label{sec:psf}

\begin{figure}
    \centering
    \includegraphics[width=\columnwidth]{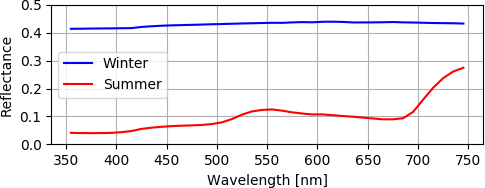}
    \caption{Reflectance spectra for the summer and winter case study. Both cases are made from a spectral mix of 60\% asphalt and 40\% grass or snow, respectively, using data from the ECOSTRESS spectral library \citep{aster}.}
    \label{fig:refl}
\end{figure}

We computed various PSFs in order to study the influence of certain parameters on light propagation, namely the AOD, the obstacle height, the ULOR of the lamps and the wavelength of the light. We also analyzed the effect of seasonal variations by simulating conditions of ground reflectance that corresponds to winter and summer using the ECOSTRESS (formerly ASTER) spectral library \citep{aster}. For summer simulations, grass represents 40\% of the spectral mix and asphalt represent 60\%. As for winter, grass is replaced with snow. The resulting spectra are shown in Figure~\ref{fig:refl}. One element of note is that the ground reflectance in winter has an almost neutral spectral effect whereas in the summer longer wavelengths are favoured. We used various obstacles height (0 m, 5 m, 7 m, 9 m, 14 m), 5 different values of the AOD (0.05, 0.10, 0.20, 0.40, 0.80), 5 different ULOR (0\%, 1\%, 5\%, 15\% and 50\%)  and wavelengths from 375 nm to 725 nm with 50 nm wide spectral bins (i.e., 7 spectral bins). In total, 182 simulations were computed in order to analyze which parameters have a significant effect on light propagation. The reference parameters chosen to compare one parameter varying at a time are presented in Table~\ref{tab:base_case}.

\begin{table}
 \centering
 \caption{Reference parameters used for comparison.}
 \label{tab:base_case}
 \begin{tabular}{lr} 
  \hline
  Parameter & Value\\
  \hline
  Wavelength & 550 nm \\
  ULOR & 5\% \\
  Obstacles height & 9 m \\
  Distance between obstacles & 20 m \\
  Obstacles filling factor & 0.8\\
  AOD@500 nm & 0.1\\
  \AA ngstr\"om exponent & 1.3 \\
  Lamp height & 7 m\\
  Relative humidity & 70\%\\
  Aerosol profile & rural\\
  Air pressure & 101.3 kPa\\
  Digital elevation model & 0 m everywhere \\
  \multirow{ 2}{*}{Ground reflectance} & 40\% Grass or snow \\
   & and 60\% Asphalt \\
  Elevation angle & 90°\\
  Azimuth Angle & 0°\\
  \hline
 \end{tabular}
\end{table}

\begin{figure*}
     \centering
     \begin{subfigure}{\columnwidth}
         \centering
         \includegraphics[width=\columnwidth]{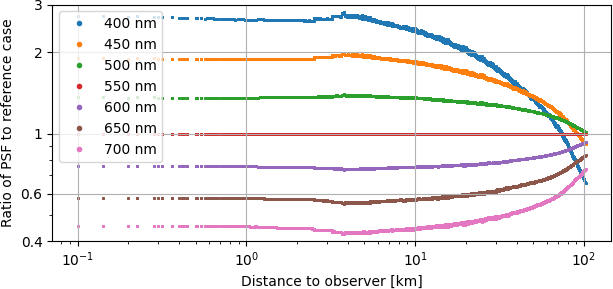}
         \caption{Winter conditions with varying wavelengths.}
         \label{fig:comparaison_hiver_wav}
     \end{subfigure}
     \begin{subfigure}{\columnwidth}
         \centering
         \includegraphics[width=\columnwidth]{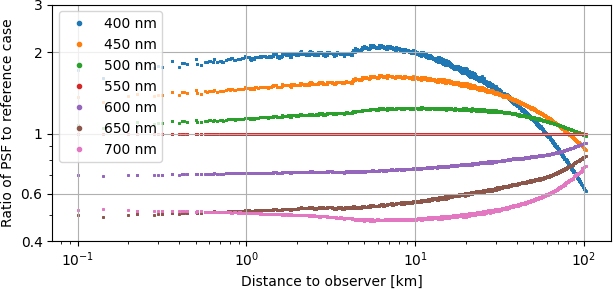}
         \caption{Summer conditions with varying wavelengths.}
         \label{fig:comparaison_ete_wav}
     \end{subfigure}
     
     \begin{subfigure}{\columnwidth}
         \centering
     \includegraphics[width=\columnwidth]{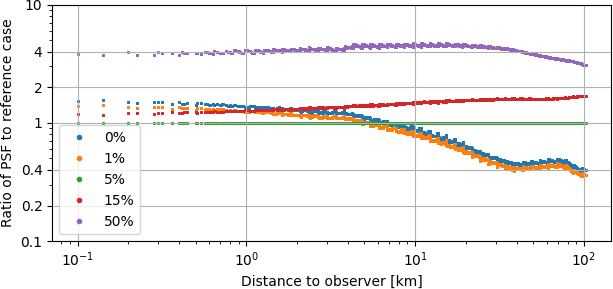}
        \caption{Winter conditions with varying ULORs.}
        \label{fig:comparaison_hiver_ULOR}
    \end{subfigure}
    \begin{subfigure}{\columnwidth}
         \centering
     \includegraphics[width=\columnwidth]{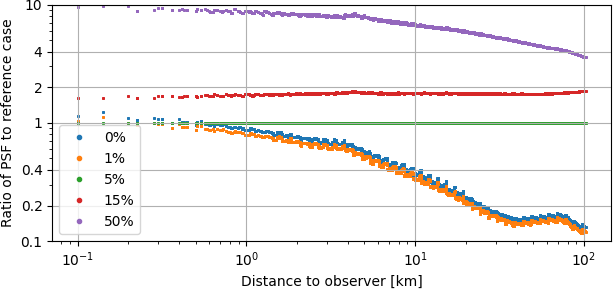}
        \caption{Summer conditions with varying ULOR.}
        \label{fig:comparaison_ete_ULOR}
    \end{subfigure}
    
     \begin{subfigure}{\columnwidth}
         \centering
 \includegraphics[width=\columnwidth]{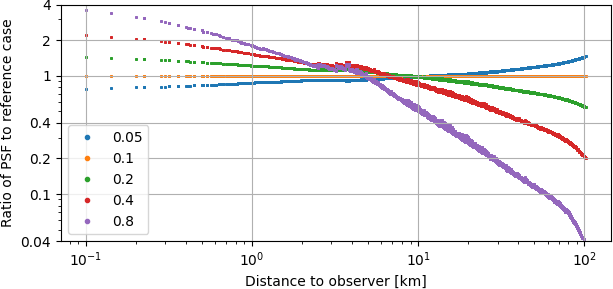}
    \caption{Winter conditions with varying AODs.}
    \label{fig:comparaison_hiver_AOD}
    \end{subfigure}
     \begin{subfigure}{\columnwidth}
         \centering
 \includegraphics[width=\columnwidth]{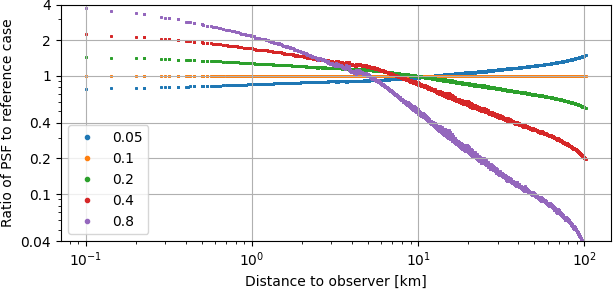}
    \caption{Summer conditions with varying AOD.}
    \label{fig:comparaison_ete_AOD}
    \end{subfigure}
    
    \begin{subfigure}{\columnwidth}
         \centering
     \includegraphics[width=\columnwidth]{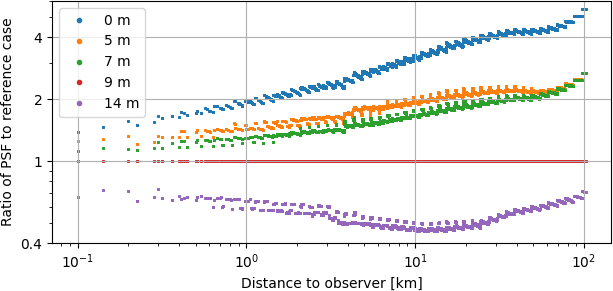}
        \caption{Winter conditions with varying obstacle heights.}
        \label{fig:comparaison_hiver_obs}
    \end{subfigure}
    \begin{subfigure}{\columnwidth}
         \centering
     \includegraphics[width=\columnwidth]{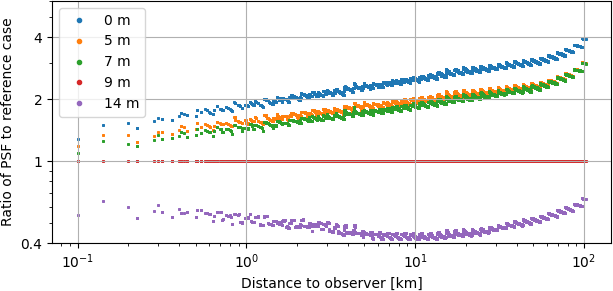}
        \caption{Summer conditions with varying obstacle heights.}
    \label{fig:comparaison_ete_obs}
    \end{subfigure}
    \caption{Point Spread Functions simulated relative to the reference parameters presented in Table \ref{tab:base_case}.}
        \label{hugefig}
\end{figure*}


To outline more clearly the change according to a given parameter in Figure \ref{hugefig}, we decided to divide any PSF by the reference parameters PSF for the given season. For that reason the reference parameters case is always equal to one. The PSFs for the reference parameters is presented in Figure \ref{fig:psf_2d_figure}. The curves order illustrated in Figure \ref{fig:comparaison_hiver_wav} is just as expected because it respects the Rayleigh's and Mie's scattering laws. Indeed, smaller wavelength causes more scattering and therefore more radiance. However, in the summer simulation there is an inversion between the curve of 700 nm and 650 nm below $\approx$800 m (See Figure \ref{fig:comparaison_ete_wav}). This inversion is caused by the spectral reflectance of grass that shows a local minimum around 680 nm followed by a rapid increase (See Figure \ref{fig:refl}). This effect is reduced at larger distances from the observer since the blocking effect of the obstacles attenuates more efficiently the reflected light than the direct light for such angles. Of course this peculiar spectral behaviour does not happen in winter when the grass is snow-covered.

In Figure \ref{fig:comparaison_hiver_ULOR}, the PSF curve of 0\% ULOR is higher than the 1\% ULOR at any distance. This may seem abnormal but it is explained by the fact that for 1\% ULOR, the light directed near the horizon is mostly blocked by obstacles with a greater height than the source but for the lamp with 0\% ULOR, the light is travelling towards the ground which reflects a part of the light towards the sky.  One difficulty to compare the different PSF for varying ULORs is that in the process, we assume the lamp flux to be constant so that lamps with lower ULORs have higher illuminance towards the ground. There is also a bump at between $\approx$40 km and $\approx$100 km. This is because of the blocking of the reflected light by obstacles. The reflected light stops reaching the line of sight which ends at 35 km above ground (maximum height of Illumina v2) at a horizontal distance of around 40 km from the source. This is why it is mainly noticeable for the 0\% and 1\% ULORs. The reflected light reaches more efficiently the overhead atmosphere at such a distance.

For the case where the AOD is varying, seen in Figure \ref{fig:comparaison_ete_AOD} and \ref{fig:comparaison_hiver_AOD}, we observe two major changes in the general tendencies of the curves. 

The first is a bump that happens between 3 km and 5 km. This bump is due to a change in the aerosol phase function that has a minimum at 115° (0° being a forward scattering direction). At larger distance the scattering angle is smaller and the phase function becomes higher. We expect this transition to happen at around 5 km assuming that most of the aerosol layer is concentrated inside the first 2 km of the atmosphere (115° corresponds to an altitude of $\approx$2.3 km at 5 km from the source). The bump is more important for an increased AOD as expected. 

The second is an inflection, observed around 80 km. It is stronger for higher AODs. This is an artefact caused by the maximum size of the 2\textsuperscript{nd} scattering radius that is fixed to 40 km into the Illumina v2 code. This feature is more important at higher AODs because that second scattering is also more important.

The inversion of the order of the curves that happen around 6 km is the effect of the attenuation mechanism increasing with distance. Close to the source, a higher AOD produce a larger radiance while it is the opposite far from the source. This is caused by the higher scattering probability, which means that for sources located far from the observer, a greater fraction of the light is scattered away before reaching the observer. However, for a source located near the observer, the higher scattering probability means that more light from the source is scattered towards the observer.

As for Figures \ref{fig:comparaison_hiver_obs} and \ref{fig:comparaison_ete_obs} with varying obstacle heights, the patterns are complex. This is caused by the obstacles height interacting with changing viewing angles with distance. Such angles are different for the direct and reflected light (i.e., ULOR and reflectance).

On all curves, the discontinuity between clusters of points that appear clearly for varying ULOR and obstacles height is due to a limitation of the Illumina v2 model. More precisely, it is caused by the obstacles that are set to be rigid transition between blocking or not blocking. This transition defines a threshold angle that interacts with the discrete step propagation of photons along the simulated line of sight. Such a numerical artefact may be removed if one allows smooth transition of the average obstacles blocking angle (some kind of blurry obstacles definition) or by reducing considerably the step length along the line of sight. The latter, of course, would increase the computing time. But it should be noted that these numerical artefacts are very small. In the plots, they are enhanced by the fact that we present ratios of PSFs. Actually, they are very hard to perceive on absolute PSFs as shown in Figure \ref{fig:psf_2d_figure}.

Overall, we clearly demonstrated that it is important to account for the parameters variations. For some curves the maximal change is more modest than others. Actually, for the varying obstacles height, the maximal change is of a factor $\approx$2. At the other extreme, varying AOD can change the PSF by a factor of $\approx$50. The variation of the PSF associated to the wavelength is a factor of  $\approx$6  while , for the ULOR, the factor is $\approx$10. The maximal factor is not happening at the same distance for all parameters. As an example, for the wavelength the variation of the PSFs is very small at large distances and maximal around 4 km. For the AOD, the variation is very small around 6 km but maximal at large distances.

\subsubsection{Fitting the PSFs}
\label{fit}

\begin{figure}
    \centering
    \includegraphics[width=\columnwidth]{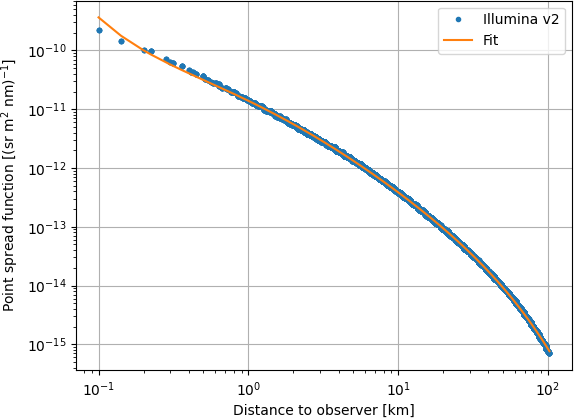}
    \caption{5th order polynomial fit of the summer reference PSF on a log-log space.}
    \label{fig:fit}
\end{figure}

In this paper we used interpolated PSF values instead of fitted functions. Such a choice render our results more accurate. However, we think that for the ease of use, finding approximate functions for the PSFs may be useful for people who are not familiar with the Illumina v2 file format. To determine the approximate functions for the PSFs, we fitted the numerical PSFs obtained with Illumina v2 by using a 5\textsuperscript{th} order polynomial in the log-log space. One such fit for  the summer reference case is presented in Figure \ref{fig:fit}.

\begin{equation}
    \log_{10}(PSF) = a + b u + c u^2 + d u^3 + e u^4 + f u^5
 \label{eq:fit}
\end{equation}

where $u\equiv\log_{10}(r)$, $r$ is the distance between the source and the observer in km and $a$, $b$, $c$, $d$, $e$, and $f$ are the fitted parameters. An extensive list of fitted parameters for the modelled cases of this paper is given in Table \ref{tab:fitparamsummer} for summer conditions and in Table \ref{tab:fitparamwinter} for winter conditions. Fit residuals show that for distances larger than 150 m, the fits do not deviate by more than 5\% from the data, if one excludes the deviations associated to the discontinuities caused by the rigid obstacles artefact discussed in Section \ref{sec:psf}. For smaller distances the error can be higher than 5\%. Therefore we recommend not using the fitted PSFs for an observer closer than 150 m from a source of light.

\begin{figure}
    \centering
    \includegraphics[width=\columnwidth]{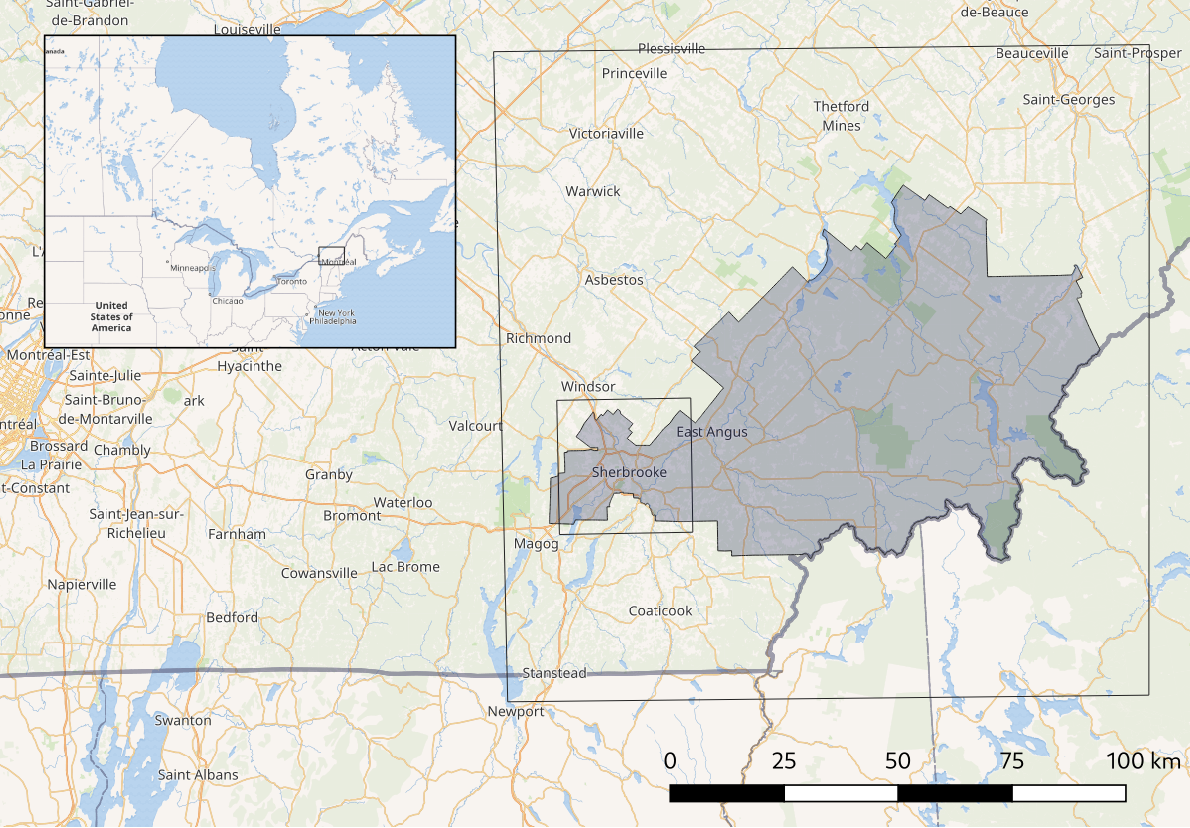}
    \caption{Simulation domain. The large square on the right represents the area for which the sky radiance maps are produced, with the smaller one in the centre, which represents the area of the city of Sherbrooke. The shaded region corresponds to the Mont-Mégantic IDSR.}
    \label{fig:domain}
\end{figure}

\subsection{Mapping the zenith luminance over the Mont-Mégantic IDSR.}
\label{sec:RICEMM}

In order to produce the ANSR maps over the Mont-Mégantic IDSR, we applied the method described above to a square area encompassing that territory (See Figure \ref{fig:domain}). Information was collected regarding the public light sources found on the territory such as their location, their height, their ULOR, their spectrum, and their luminous flux. We determined characteristics for about 15,000 public fixtures on the modelled domain. It is important to acknowledge that we do not consider the effect of any lamp outside the limits of the IDSR nor the private sources or vehicle headlights inside the limits. We know from previous work with Illumina that the effect of neglecting small sources outside the domain is not very important for the zenith radiance \citep{HAWAII} but neglecting private sources and vehicles may be a major drawback. Private sources can represent a substantial fraction of the total ALAN and consequently of the total ANSR. As an example, \citet{simoneau2020multispectral} estimated the private residential sources to contribute up to 15\% of the total zenith ANSR at Asiago observatory. \citet{sanchezdemiguel2015variacion} estimated the private sources to contribute up to 45\% in Madrid, Spain, and \citet{bara2019estimating} obtained private contributions before midnight up to 22\% and 42\% for two Spanish cities while the vehicles contribute to less than 1\%. The private contribution is reducing with time of the night. \citet{barentine2020recovering}, estimated the contribution of public sources to the zenith ANSR to be ranging between 3\% to 14\% depending on the method used (i.e., private between $\sim$ 86\% and 97\%). More recently \citet{kybapublic} showed that public lighting contributes to 18\% of the radiance detected by VIIRS-DNB for Tucson, USA.  According to such results, we may expect a huge discrepancy between our modelled results and the SQC measurements. 

Illumina v2 input data tools were used to split the inventory map of lamps into different maps of luminous flux for each spectral bin and for each ULOR. In all cases we assumed the lamp height and the obstacles properties to be the same everywhere on the modelling domain. Their reference values are given as the reference parameters in Table \ref{tab:base_case}. 

Once the PSFs were determined for all cases, we performed the convolutions over the corresponding luminous flux maps for all ULORs and spectral bins. Since the real sky radiance is produced by the combination of all lights, we sum together all the convoluted maps per ULOR to produce one zenith ANSR map per spectral bin.

\begin{figure*}
\begin{multicols}{2}
 \includegraphics[width=\columnwidth]{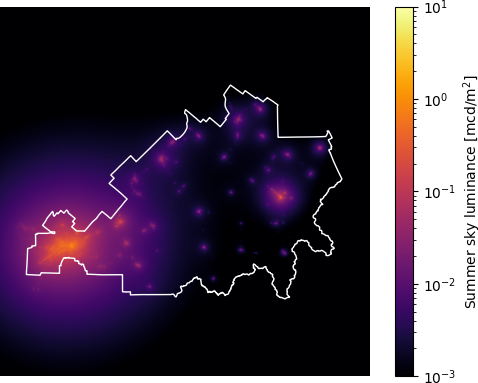}
    \caption{Mont-Mégantic IDSR artificial zenith sky luminance computed using the base PSF case presented in Table \ref{tab:base_case} in summer conditions. The limits of the IDSR are shown in white.}
    \label{fig:carte_ete}
%
 \includegraphics[width=\columnwidth]{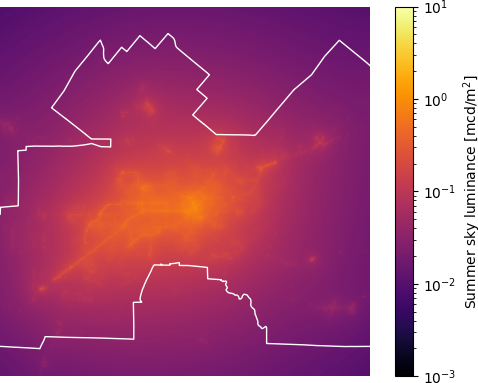}
    \caption{City of Sherbrooke artificial zenith night sky luminance computed using the base PSF case presented in Table \ref{tab:base_case} in summer conditions. The limits of the IDSR are shown in white.}
    \label{fig:carte_sherbrooke_ete}

 \includegraphics[width=\columnwidth]{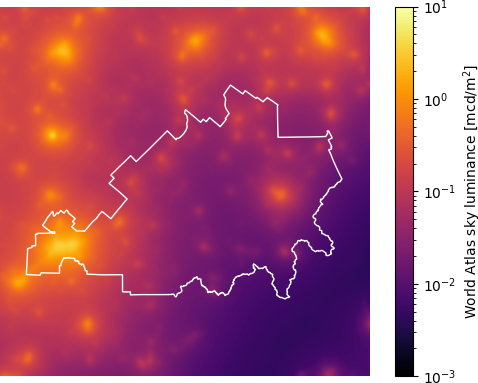}
    \caption{Artificial zenith sky luminance from the New World Atlas of Artificial Sky brightness over the Mont-Mégantic IDSR. The limits of the IDSR are shown in white.}
    \label{atlasricemm}
%
 \includegraphics[width=\columnwidth]{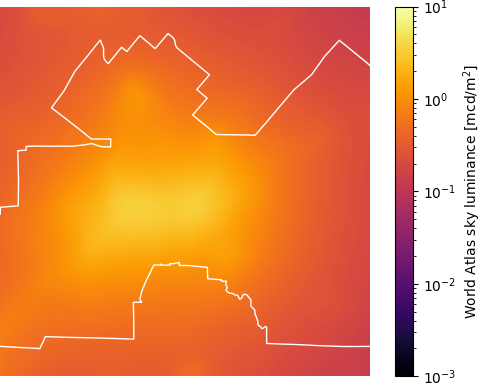}
    \caption{Artificial zenith sky luminance of the New World Atlas of Artificial Sky brightness over the Sherbrooke. The limits of the IDSR are shown in white.}
    \label{atlassherbrooke}
\end{multicols}
\end{figure*}

The resultant zenith ANSR spectral maps $L_{e,\lambda}$ can be integrated according to any spectral sensitivity depending on the desired application. It can either be the photosynthesis action spectrum for ecological studies on flora or the scotopic spectral sensitivity which correspond low light human eye sensitivity \citep{aube2013evaluating} for when the interest is in stargazing applications. It is also possible to use the Sky Quality Meter (SQM) \citep{cinzano2005night} sensitivity spectra to compare SQM measurements with the model predictions. In this study, we decided to convert the spectral radiances $L_{e,\lambda}$ obtained from the convolutions to luminance values $L$ in units of Cd/m$^2$. This is done using

\begin{equation}
    L = K_m \int V(\lambda) L_{e,\lambda}~\mathrm{d}\lambda
 \label{eq:final_map_integral}
\end{equation}

where $K_m$ is the maximum spectral luminous efficacy and has a value of 683 lum/W and $V(\lambda)$ is the photopic luminosity function. The spectral sensitivity used in this case is the photopic sensitivity. We end up with the map presented in Figure~\ref{fig:carte_ete}. Figure \ref{fig:carte_sherbrooke_ete} shows a close-up on the city of Sherbrooke.

The artificial zenith sky luminance shown in Figures \ref{fig:carte_ete} and \ref{fig:carte_sherbrooke_ete} can be compared with the artificial part of the NWA (Figures \ref{atlasricemm} and \ref{atlassherbrooke}). That comparison shows that our modelled luminances are generally underestimated compared to the NWA. We postulate that this is because our light fixtures inventory is not complete (missing private sources). 

Figure \ref{fig:carte_sherbrooke_ete} show that our computations find high frequency spatial structures to the artificial zenith luminance map with peaks corresponding to the main streets of the city. Such high frequencies are not visible in the NWA (see Figure \ref{atlassherbrooke}). For the NWA, the resolution of the model used was not high enough to reveal such a fine structure. The high frequency structure may likely be less visible when considering all the sources (private and public) but we do not have the data to validate this hypothesis.

\subsection{Comparison with Sky Quality Camera measurements}
\label{sec:sqc}

\begin{figure}
    \centering
    \includegraphics[width=\columnwidth]{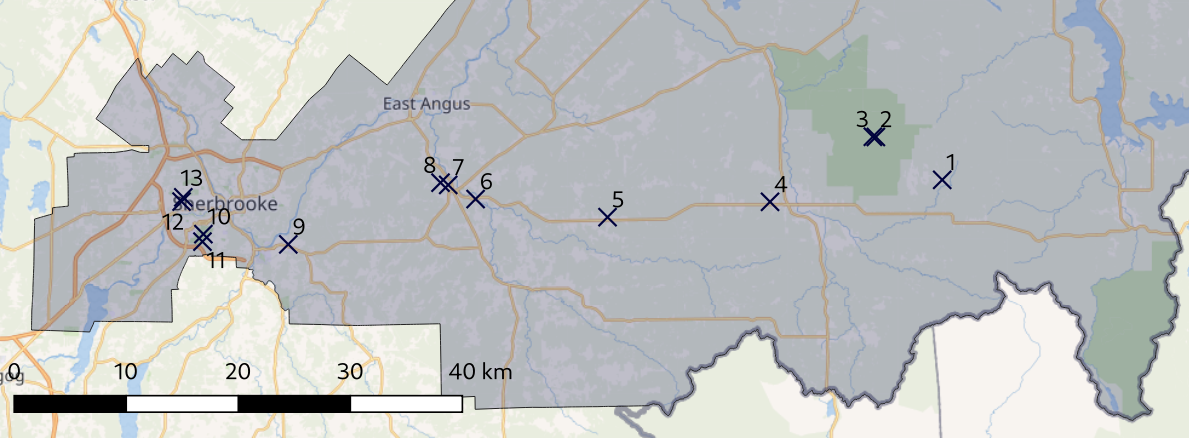}
    \caption{Position of the measurements made with the SQC over the Mont-Mégantic IDSR (shaded region). The numbers refer to the numbers of column 1 in Table \ref{tab:compdata}.}
    \label{fig:domainpoints}
\end{figure}

To properly compare our model results to SQC measurements, we first need to remove the natural sky luminance value of 0.174 mcd/m$^2$ \citep{puschnig2014night} from the measured luminance. The first 11 zenithal luminances of Table \ref{tab:compdata} and Figure \ref{fig:domainpoints} were extracted from SQC images taken on 5 different nights, between May 7, 2019, and April 20, 2020, and representing a West-to-East transect from Sherbrooke to Mont-Mégantic. Measurements \#12 and \#13 were taken on August 27, 2020. All images were taken without snow cover. The images were selected according to the visual quality of the night sky and the absence of the Milky Way directly at zenith. While the Milky Way plane was at least 40° from the zenith on most of the nights, it was between 15° and 20° for the three measurements closest to Mont Bellevue in Sherbrooke (\#10 \& \#11), where its influence in the luminance of the night sky will have relatively less impact because of the higher level of artificial sky luminance. For the last two data of the table, the Milky Way plane was only 10-11° from zenith. Measurements were taken between 11:30 p.m. and 2:00 a.m. local time, after astronomical twilight, and in the absence of the Moon and clouds.

\begin{table*}
    \centering
    \caption{Comparison of the modelled artificial luminances with the observations made with the SQC during a transect spanning from Mont-Mégantic to Mont Bellevue (Sherbrooke City) from which we removed an estimate of the natural luminance of 0.174 mcd m$^{-2}$. All data points correspond to summer conditions. 'Convolution' corresponds to the present work.}
    \label{tab:compdata}
    \begin{tabular}{cccccccccc}
\hline
        Site & Latitude & Longitude & Elevation & SQC & New World Atlas &  Convolution & Uniform & Optimal  \\ 
         &  &  &  & (artificial) & (artificial) &  &
        Illumina v2 & Illumina v2 \\         
        No & deg & deg & m & mcd m$^{-2}$ & mcd m$^{-2}$ & mcd m$^{-2}$ & mcd m$^{-2}$ & mcd m$^{-2}$ \\        
        \hline
         1 & 45.4214 & -71.0740 &  521 & 0.002 & 0.017 & 0.001 & 0.002 & 0.005 \\
         2 & 45.4555 & -71.1494 & 1090 & 0.012 & 0.017 & 0.001 & 0.002 & 0.003 \\
         3 & 45.4557 & -71.1528 & 1111 & 0.019 & 0.017 & 0.001 & 0.004 & 0.006 \\
         4 & 45.4038 & -71.2697 &  457 & 0.036 & 0.040 & 0.004 & 0.004 & 0.011 \\
         5 & 45.3916 & -71.4543 &  383 & 0.060 & 0.041 & 0.004 & 0.007 & 0.011 \\
         6 & 45.4060 & -71.6042 &  207 & 0.130 & 0.129 & 0.014 & 0.020 & 0.036 \\
         7 & 45.4173 & -71.6353 &  240 & 0.243 & 0.230 & 0.033 & 0.040 & 0.100 \\
         8 & 45.4191 & -71.6440 &  241 & 0.167 & 0.207 & 0.021 & 0.030 & 0.059 \\
         9 & 45.3699 & -71.8170 &  192 & 0.510 & 0.652 & 0.066 & 0.133 & 0.223 \\
        10 & 45.3781 & -71.9134 &  323 & 1,556 & 2,060 & 0.232 & 0.262 & 0.554 \\
        11 & 45.3721 & -71.9145 &  349 & 1,175 & 1,710 & 0.183 & 0.218 & 0.445 \\
        12 & 45.4040 & -71.9382 &  211 & 2,230 & 3,384 & 0.487 & 0.504 & 1,228 \\
        13 & 45.4074 & -71.9362 &  241 & 1,827 & 3,144 & 0.439 & 0.455 & 1,062 \\
        \hline
    \end{tabular}
\end{table*}

\begin{figure}
 \includegraphics[width=\columnwidth]{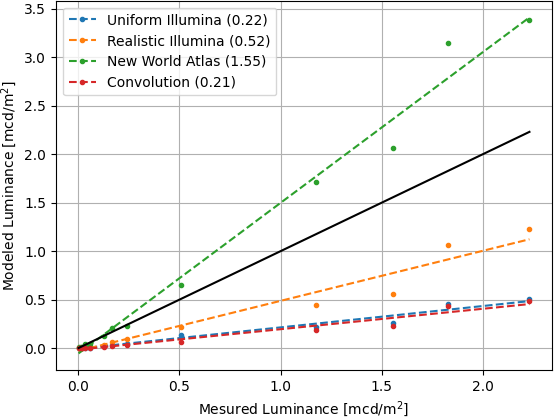}
    \caption{Scatter plot between the zenith artificial luminance measured with the SQC (SQC measurement minus 0.174 mcd/m$^2$) and the modelled artificial luminances. The black line is a perfect 1:1 relation, and the value in parentheses is the slope of a linear regression. 'Uniform Illumina' is a modelling with Illumina v2 assuming no topography, 'Realistic Illumina' includes the topography and 'Convolution' is the FFT method developed in this study.}
    \label{fig:scatter}
\end{figure}

\begin{figure}
    \centering
    \includegraphics[width=\columnwidth]{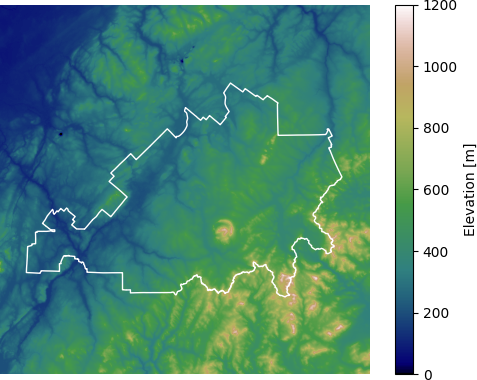}
    \caption{Topography of the simulation domain. The limits of the IDSR are shown in white.}
    \label{fig:srtm}
\end{figure}

Figure \ref{fig:scatter} and Table \ref{tab:compdata} show the comparison between the different models along with the SQC measurements (natural luminance removed) for the 13 measuring points. For the convolution method and Illumina models, the reference parameters in summer were used except for the lamps spectra, luminous flux and ULOR that were determined from the light fixtures inventory. The 'Uniform Illumina' model uses the same input parameters as the 'convolution' method in order to compare the validity of the approach. We therefore expect the two to produce very similar results, and it is what we obtained. We use a linear regression on the data to obtain a relationship between the modelled luminances and the measured ones. That regression will be less sensitive to the darkest values, which are greatly affected by small changes in the natural sky luminance value used. The methods have slopes with the measured luminance of 0.22 and 0.21 respectively. The latter result indicates that, at least for the case of Mont-Mégantic IDSR, the use of the convolution method is a valuable alternative to the computationally demanding Illumina v2 model when no georeferenced information about the obstacles and topography are available. The 'Realistic Illumina' model adds the domain topography to study the impacts of flattening the simulation domain on the accuracy of the results. The topography of the simulated domain is shown in Figure~\ref{fig:srtm}. When comparing the 'convolution' method with the 'Realistic Illumina' method, we find important discrepancies. The slope of Figure \ref{fig:scatter} being 0.52 for the 'Realistic Illumina' while it is 0.21 for the 'convolution' method. In other words, in the studied case, the convolution method predicts less than half (0.4) of the artificial zenith luminance that a more complete Illumina v2 simulation would do. But the Mont-Mégantic IDSR territory shows relatively high topography variations (see Figure \ref{fig:srtm}).

Based on a comparison of the 'Realistic Illumina' model with the Sky Quality Camera luminance measurements, we determined that over the Mont-Mégantic IDSR, roughly 48 percent of the artificial sky luminance towards zenith potentially could be associated to the missing private sources. We obtain this number from the slope of Figure \ref{fig:scatter} (i.e. $1-0.52=0.48$). Such a result seems to be coherent with previous estimations made by many authors in a variety of sites and with a variety of methods \citep{sanchezdemiguel2015variacion,bara2019estimating,barentine2020recovering,kybapublic,simoneau2020multispectral}.
  
Basically, we find that a convolution based method to determine the zenith sky luminance is not valid for non-flat territories. This greatly reduces the potential of such a method, potentially restricting it to flat lands like the Netherlands or any plain or plateau as long as the main contributing sources are inside that region. 

Comparing the values provided by the NWA with the SQC measurements shows that the NWA seems to be overestimating sky luminance by 55\%. This can be explained by a variety of reasons, including a change in atmospheric and reflectance conditions between the gathering of both datasets. Moreover, NWA averaged 6 months of VIIRS-DNB data from May to December 2014. We know that there was snow cover for at least two of these months. This effect alone can explain the NWA overestimation. Nevertheless the NWA assumes uniform properties of light sources for the whole world and tries to compensate for the blue-light blindness of the VIIRS sensor \citep{falchi2016} from which it obtained the data used for the model. It is therefore probable that the NWA also overestimate the sky luminance of Mont-Mégantic IDSR due to the abundance of PC Amber LEDs and full cutoff lights on the territory. Finally, the VIIRS sensor is also sensitive to the near infrared emitted by high pressure sodium lamps whereas the SQC is not. Because of the abundance of those lights in the region it could also contribute to the observed difference.

In order to validate that the fine structure we can see in the luminance maps obtained from the convolution method (Figures \ref{fig:carte_ete} and \ref{fig:carte_sherbrooke_ete}) that is absent from the NWA maps (Figures \ref{atlasricemm} and \ref{atlassherbrooke}) is real, we can compare the measurements from points 12 and 13, located within 400m of each other inside the city of Sherbrooke. Point 12 is located on a main avenue whereas point 13 is in a nearby parc. We notice that they exhibit a much greater relative difference between their luminance in the SQC data ($\approx17\%$) than in the NWA ($\approx$4\%). The relative difference predicted by the convolution method is $\approx$12\%, which is much closer to the measured difference. This suggests that the fine structure we see on Figures~\ref{fig:carte_ete} and \ref{fig:carte_sherbrooke_ete} might be real, although a more complete study would be needed.

\section{Conclusions}

We observe that the variations in the parameters considered in that study have strong impacts on the PSFs. This indicates that the ULOR and wavelength emitted by a light source along with the AOD and obstacle height of a given environment are key parameters to define the PSFs used to compute the sky radiance/luminance. We actually found that, in the limits of the tested parameters variations, the AOD has the strongest impact on the PSF ($ \approx50\times$), followed by the ULOR ($\approx10\times$), the wavelength ($\approx6\times$) and the obstacles height ($\approx2\times$).  The PSFs are also significantly impacted by the season, but it is mostly a normalization change as the shape of the curves stays approximately the same.

We tested if the use of convolutions of PSFs derived with the light pollution model Illumina v2 (without topography) can be used to produce sky radiance maps over large territories. When it comes to the application of the PSFs to the territory of Mont-Mégantic IDSR, we can observe that the city roads are distinguishable while they are not in the New world atlas of artificial sky brightness. This seems coherent with the SQC observations for two validation points located within 400m of each other in the city of Sherbrooke, one on a main avenue and the other in a park. This shows that the method we used produces maps with higher resolution. We also find that a convolution based method to determine the artificial zenith sky luminance is not valid for non-flat territories. This greatly reduces the potential of the method. But for a non-flat territory, like the Mount-Mégantic IDSR, we found the convolution method to be underestimated by a factor of $\approx2.5$. One possible idea to explore would be to find typical correction factors to the 'convolution' method according to the typical topography of the modelled territory. In our case a correction factor of $\approx2.5\times$ compared with the 'Realistic Illumina' simulation. At the end, if we apply such correction factor and if we use a complete light inventory (private and public) we should have good 'convolution' method predictions of the ANSR. Further studies must be done to determine how the correction factor depends on the statistical properties of the topography features.

Finding private sources inventory can be done using airborne observations with drones \citep{fiorentin2018minlu,bouroussis2020assessment}, aircraft \citep{kuechly2012aerial,hale2013mapping}, or stratospheric balloon flights \citep{ocana2016low}. Our research group is developing a stratospheric balloon observation system named High Altitude Balloon Light at Night (HABLAN, \citealt{hablan}) in partnership with the Canadian Space Agency to address that issue. The latter project aims to determine the source’s positions, spectra and angular emission functions from which ULOR may be determined. 

\section*{Acknowledgements}

We applied the sequence-determines-credit approach \citep{tscharntke2007author} for the sequence of authors which is in order of decreasing contributions. Some computations were carried out on the Mammouth Serial II cluster managed by Calcul Québec and Compute Canada. The operation of this supercomputer is funded by the Canada Foundation for Innovation (CFI), NanoQuébec, Réseau de Médecine Génétique Appliquée, and the Fonds de recherche du Québec -- Nature et technologies (FRQNT).  M.\,A. and A.\,S.~thanks the FRQNT for financial support through the Research program for college researchers. J. R., J.\,L. and F.\,L.~thanks the Pôle régional en enseignement supérieur de l'Estrie (PRESE). Measurements with the SQC were made possible thanks to the contribution of the Mont-Mégantic National Park. Mont-Mégantic National Park and Hydro-Sherbrooke provided part of the sources inventory.

\section*{Data availability}

The data underlying this article will be shared on reasonable request to the corresponding author.




\bibliographystyle{mn2e}
\bibliography{references}




\onecolumn
\appendix
\section{Fit parameters of the PSFs}
\small
\begin{longtable}{ccccccccccc}
    \caption{Fitted parameters for the various summer cases modelled in this paper.}
    \\
    \hline
    No.  & Obst. height & ULOR & AOD & Wavelength & a & b & c & d & e & f \\
     &   m & \% &  & nm &  & &  &  & &    \\   
     \hline
\endfirsthead
\multicolumn{11}{c}{\tablename\ \thetable{} -- continued from previous page} \\
\hline
    No.  & Obst. height & ULOR & AOD & Wavelength & a & b & c & d & e & f \\
     &   m & \% &  & nm &  & &  &  & &    \\
     \hline
\endhead
\endlastfoot
1  & 0  & 5  & 0.1  & 400 & -10.322 & -1.088 & -0.257 & -0.213 & 0.114 & -0.048 \\
2  & 0  & 5  & 0.1  & 450 & -10.427 & -1.096 & -0.239 & -0.176 & 0.114 & -0.048 \\
3  & 0  & 5  & 0.1  & 500 & -10.534 & -1.116 & -0.220 & -0.140 & 0.093 & -0.041 \\
4  & 0  & 5  & 0.1  & 550 & -10.581 & -1.144 & -0.214 & -0.124 & 0.089 & -0.038 \\
5  & 0  & 5  & 0.1  & 600 & -10.727 & -1.146 & -0.205 & -0.098 & 0.071 & -0.033 \\
6  & 0  & 5  & 0.1  & 650 & -10.864 & -1.147 & -0.194 & -0.073 & 0.048 & -0.026 \\
7  & 0  & 5  & 0.1  & 700 & -10.863 & -1.179 & -0.193 & -0.071 & 0.048 & -0.024 \\
8  & 5  & 5  & 0.1  & 400 & -10.366 & -1.089 & -0.255 & -0.223 & 0.121 & -0.050 \\
9  & 5  & 5  & 0.1  & 450 & -10.482 & -1.100 & -0.238 & -0.191 & 0.125 & -0.050 \\
10 & 5  & 5  & 0.1  & 500 & -10.597 & -1.121 & -0.223 & -0.159 & 0.112 & -0.045 \\
11 & 5  & 5  & 0.1  & 550 & -10.661 & -1.161 & -0.214 & -0.143 & 0.104 & -0.041 \\
12 & 5  & 5  & 0.1  & 600 & -10.802 & -1.167 & -0.201 & -0.106 & 0.076 & -0.033 \\
13 & 5  & 5  & 0.1  & 650 & -10.935 & -1.161 & -0.193 & -0.087 & 0.061 & -0.028 \\
14 & 5  & 5  & 0.1  & 700 & -10.952 & -1.212 & -0.192 & -0.075 & 0.053 & -0.025 \\
15 & 7  & 5  & 0.1  & 400 & -10.380 & -1.099 & -0.252 & -0.202 & 0.103 & -0.045 \\
16 & 7  & 5  & 0.1  & 450 & -10.501 & -1.110 & -0.234 & -0.171 & 0.109 & -0.046 \\
17 & 7  & 5  & 0.1  & 500 & -10.618 & -1.127 & -0.223 & -0.146 & 0.104 & -0.044 \\
18 & 7  & 5  & 0.1  & 550 & -10.688 & -1.170 & -0.216 & -0.126 & 0.098 & -0.041 \\
19 & 7  & 5  & 0.1  & 600 & -10.827 & -1.169 & -0.204 & -0.099 & 0.077 & -0.034 \\
20 & 7  & 5  & 0.1  & 650 & -10.958 & -1.164 & -0.197 & -0.076 & 0.061 & -0.029 \\
21 & 7  & 5  & 0.1  & 700 & -10.982 & -1.215 & -0.192 & -0.069 & 0.056 & -0.027 \\
22 & 9  & 0  & 0.1  & 400 & -10.925 & -1.110 & -0.629 & -0.718 & 0.728 & -0.206 \\
23 & 9  & 0  & 0.1  & 450 & -10.918 & -1.140 & -0.566 & -0.677 & 0.634 & -0.168 \\
24 & 9  & 0  & 0.1  & 500 & -10.966 & -1.202 & -0.495 & -0.606 & 0.508 & -0.122 \\
25 & 9  & 0  & 0.1  & 550 & -10.923 & -1.268 & -0.425 & -0.533 & 0.381 & -0.079 \\
26 & 9  & 0  & 0.1  & 600 & -11.098 & -1.304 & -0.387 & -0.506 & 0.317 & -0.053 \\
27 & 9  & 0  & 0.1  & 650 & -11.273 & -1.344 & -0.358 & -0.468 & 0.255 & -0.031 \\
28 & 9  & 0  & 0.1  & 700 & -11.186 & -1.409 & -0.307 & -0.387 & 0.147 & 0.002  \\
29 & 9  & 1  & 0.1  & 400 & -10.966 & -1.111 & -0.622 & -0.715 & 0.721 & -0.204 \\
30 & 9  & 1  & 0.1  & 450 & -10.961 & -1.145 & -0.559 & -0.667 & 0.625 & -0.166 \\
31 & 9  & 1  & 0.1  & 500 & -11.008 & -1.206 & -0.492 & -0.596 & 0.503 & -0.122 \\
32 & 9  & 1  & 0.1  & 550 & -10.965 & -1.266 & -0.427 & -0.533 & 0.386 & -0.081 \\
33 & 9  & 1  & 0.1  & 600 & -11.139 & -1.304 & -0.389 & -0.502 & 0.319 & -0.055 \\
34 & 9  & 1  & 0.1  & 650 & -11.312 & -1.343 & -0.370 & -0.462 & 0.268 & -0.038 \\
35 & 9  & 1  & 0.1  & 700 & -11.227 & -1.404 & -0.319 & -0.389 & 0.164 & -0.005 \\
36 & 9  & 5  & 0.05 & 400 & -10.628 & -1.152 & -0.190 & -0.177 & 0.119 & -0.060 \\
37 & 9  & 5  & 0.05 & 450 & -10.748 & -1.162 & -0.208 & -0.159 & 0.144 & -0.065 \\
38 & 9  & 5  & 0.05 & 500 & -10.866 & -1.184 & -0.211 & -0.135 & 0.134 & -0.060 \\
39 & 9  & 5  & 0.05 & 550 & -10.924 & -1.220 & -0.240 & -0.127 & 0.145 & -0.059 \\
40 & 9  & 5  & 0.05 & 600 & -11.082 & -1.222 & -0.221 & -0.108 & 0.119 & -0.050 \\
41 & 9  & 5  & 0.05 & 650 & -11.231 & -1.224 & -0.204 & -0.088 & 0.093 & -0.041 \\
42 & 9  & 5  & 0.05 & 700 & -11.242 & -1.259 & -0.231 & -0.103 & 0.109 & -0.041 \\
43 & 9  & 5  & 0.1  & 400 & -10.589 & -1.150 & -0.215 & -0.273 & 0.177 & -0.068 \\
44 & 9  & 5  & 0.1  & 450 & -10.696 & -1.169 & -0.225 & -0.231 & 0.183 & -0.071 \\
45 & 9  & 5  & 0.1  & 500 & -10.803 & -1.189 & -0.241 & -0.205 & 0.190 & -0.073 \\
46 & 9  & 5  & 0.1  & 550 & -10.856 & -1.235 & -0.254 & -0.184 & 0.184 & -0.069 \\
47 & 9  & 5  & 0.1  & 600 & -11.001 & -1.241 & -0.244 & -0.156 & 0.162 & -0.062 \\
48 & 9  & 5  & 0.1  & 650 & -11.139 & -1.245 & -0.233 & -0.129 & 0.136 & -0.055 \\
49 & 9  & 5  & 0.1  & 700 & -11.146 & -1.293 & -0.249 & -0.126 & 0.138 & -0.052 \\
50 & 9  & 5  & 0.2  & 400 & -10.533 & -1.140 & -0.295 & -0.452 & 0.314 & -0.092 \\
51 & 9  & 5  & 0.2  & 450 & -10.623 & -1.161 & -0.292 & -0.392 & 0.302 & -0.093 \\
52 & 9  & 5  & 0.2  & 500 & -10.714 & -1.202 & -0.290 & -0.322 & 0.270 & -0.088 \\
53 & 9  & 5  & 0.2  & 550 & -10.757 & -1.250 & -0.303 & -0.288 & 0.265 & -0.087 \\
54 & 9  & 5  & 0.2  & 600 & -10.887 & -1.267 & -0.284 & -0.236 & 0.219 & -0.076 \\
55 & 9  & 5  & 0.2  & 650 & -11.009 & -1.274 & -0.272 & -0.198 & 0.185 & -0.067 \\
56 & 9  & 5  & 0.2  & 700 & -11.012 & -1.337 & -0.273 & -0.169 & 0.164 & -0.060 \\
57 & 9  & 5  & 0.4  & 400 & -10.475 & -1.138 & -0.491 & -0.741 & 0.595 & -0.154 \\
58 & 9  & 5  & 0.4  & 450 & -10.541 & -1.161 & -0.454 & -0.642 & 0.534 & -0.143 \\
59 & 9  & 5  & 0.4  & 500 & -10.613 & -1.206 & -0.420 & -0.540 & 0.458 & -0.127 \\
60 & 9  & 5  & 0.4  & 550 & -10.643 & -1.270 & -0.399 & -0.459 & 0.400 & -0.114 \\
61 & 9  & 5  & 0.4  & 600 & -10.754 & -1.295 & -0.372 & -0.380 & 0.330 & -0.098 \\
62 & 9  & 5  & 0.4  & 650 & -10.860 & -1.308 & -0.350 & -0.320 & 0.273 & -0.084 \\
63 & 9  & 5  & 0.4  & 700 & -10.854 & -1.377 & -0.336 & -0.268 & 0.230 & -0.073 \\
64 & 9  & 5  & 0.8  & 400 & -10.452 & -1.214 & -0.938 & -1.120 & 1.118 & -0.290 \\
65 & 9  & 5  & 0.8  & 450 & -10.489 & -1.217 & -0.809 & -0.983 & 0.956 & -0.249 \\
66 & 9  & 5  & 0.8  & 500 & -10.535 & -1.257 & -0.715 & -0.837 & 0.803 & -0.211 \\
67 & 9  & 5  & 0.8  & 550 & -10.550 & -1.317 & -0.639 & -0.716 & 0.677 & -0.180 \\
68 & 9  & 5  & 0.8  & 600 & -10.635 & -1.350 & -0.583 & -0.596 & 0.555 & -0.150 \\
69 & 9  & 5  & 0.8  & 650 & -10.720 & -1.370 & -0.535 & -0.500 & 0.453 & -0.125 \\
70 & 9  & 5  & 0.8  & 700 & -10.705 & -1.438 & -0.488 & -0.420 & 0.371 & -0.105 \\
71 & 9  & 15 & 0.1  & 400 & -10.264 & -1.128 & -0.271 & -0.318 & 0.226 & -0.078 \\
72 & 9  & 15 & 0.1  & 450 & -10.401 & -1.141 & -0.262 & -0.271 & 0.215 & -0.075 \\
73 & 9  & 15 & 0.1  & 500 & -10.526 & -1.162 & -0.261 & -0.234 & 0.202 & -0.071 \\
74 & 9  & 15 & 0.1  & 550 & -10.619 & -1.199 & -0.261 & -0.202 & 0.184 & -0.065 \\
75 & 9  & 15 & 0.1  & 600 & -10.752 & -1.214 & -0.249 & -0.167 & 0.153 & -0.056 \\
76 & 9  & 15 & 0.1  & 650 & -10.877 & -1.224 & -0.238 & -0.140 & 0.127 & -0.048 \\
77 & 9  & 15 & 0.1  & 700 & -10.926 & -1.258 & -0.244 & -0.132 & 0.123 & -0.045 \\
78 & 9  & 50 & 0.1  & 400 & -9.500  & -1.131 & -0.369 & -0.472 & 0.408 & -0.128 \\
79 & 9  & 50 & 0.1  & 450 & -9.651  & -1.160 & -0.349 & -0.398 & 0.360 & -0.115 \\
80 & 9  & 50 & 0.1  & 500 & -9.792  & -1.208 & -0.320 & -0.320 & 0.288 & -0.094 \\
81 & 9  & 50 & 0.1  & 550 & -9.921  & -1.244 & -0.307 & -0.269 & 0.243 & -0.081 \\
82 & 9  & 50 & 0.1  & 600 & -10.044 & -1.279 & -0.287 & -0.221 & 0.192 & -0.066 \\
83 & 9  & 50 & 0.1  & 650 & -10.158 & -1.304 & -0.277 & -0.189 & 0.159 & -0.056 \\
84 & 9  & 50 & 0.1  & 700 & -10.254 & -1.329 & -0.273 & -0.162 & 0.133 & -0.048 \\
85 & 14 & 5  & 0.1  & 400 & -10.938 & -1.197 & -0.287 & -0.225 & 0.183 & -0.072 \\
86 & 14 & 5  & 0.1  & 450 & -11.016 & -1.237 & -0.292 & -0.183 & 0.198 & -0.078 \\
87 & 14 & 5  & 0.1  & 500 & -11.106 & -1.279 & -0.291 & -0.142 & 0.189 & -0.075 \\
88 & 14 & 5  & 0.1  & 550 & -11.129 & -1.333 & -0.323 & -0.126 & 0.214 & -0.082 \\
89 & 14 & 5  & 0.1  & 600 & -11.281 & -1.339 & -0.295 & -0.094 & 0.178 & -0.071 \\
90 & 14 & 5  & 0.1  & 650 & -11.432 & -1.347 & -0.257 & -0.054 & 0.127 & -0.056 \\
91 & 14 & 5  & 0.1  & 700 & -11.409 & -1.398 & -0.301 & -0.059 & 0.164 & -0.066 \\
\hline
\label{tab:fitparamsummer}
\end{longtable}

\begin{longtable}{ccccccccccc}
    \caption{Fitted parameters for the various winter cases modelled in this paper.}
    \\
    \hline
    No.  & Obst. height & ULOR & AOD & Wavelength & a & b & c & d & e & f \\
     &   m & \% &  & nm &  & &  &  & &    \\   
     \hline
\endfirsthead
\multicolumn{11}{c}{\tablename\ \thetable{} -- continued from previous page} \\
\hline
    No.  & Obst. height & ULOR & AOD & Wavelength & a & b & c & d & e & f \\
     &   m & \% &  & nm &  & &  &  & &    \\
     \hline
\endhead
\endlastfoot

1  & 0  & 5  & 0.1  & 400 & -9.779  & -1.136 & -0.303 & -0.324 & 0.240 & -0.081 \\
2  & 0  & 5  & 0.1  & 450 & -9.927  & -1.156 & -0.275 & -0.251 & 0.197 & -0.069 \\
3  & 0  & 5  & 0.1  & 500 & -10.069 & -1.177 & -0.251 & -0.200 & 0.158 & -0.056 \\
4  & 0  & 5  & 0.1  & 550 & -10.202 & -1.199 & -0.232 & -0.157 & 0.122 & -0.046 \\
5  & 0  & 5  & 0.1  & 600 & -10.321 & -1.216 & -0.221 & -0.126 & 0.097 & -0.038 \\
6  & 0  & 5  & 0.1  & 650 & -10.432 & -1.225 & -0.218 & -0.109 & 0.087 & -0.034 \\
7  & 0  & 5  & 0.1  & 700 & -10.535 & -1.238 & -0.208 & -0.085 & 0.066 & -0.028 \\
8  & 5  & 5  & 0.1  & 400 & -9.908  & -1.134 & -0.354 & -0.417 & 0.343 & -0.108 \\
9  & 5  & 5  & 0.1  & 450 & -10.055 & -1.166 & -0.313 & -0.337 & 0.275 & -0.087 \\
10 & 5  & 5  & 0.1  & 500 & -10.196 & -1.204 & -0.285 & -0.270 & 0.218 & -0.069 \\
11 & 5  & 5  & 0.1  & 550 & -10.327 & -1.238 & -0.262 & -0.215 & 0.168 & -0.053 \\
12 & 5  & 5  & 0.1  & 600 & -10.450 & -1.270 & -0.238 & -0.167 & 0.120 & -0.039 \\
13 & 5  & 5  & 0.1  & 650 & -10.563 & -1.286 & -0.234 & -0.142 & 0.103 & -0.034 \\
14 & 5  & 5  & 0.1  & 700 & -10.665 & -1.303 & -0.229 & -0.114 & 0.086 & -0.029 \\
15 & 7  & 5  & 0.1  & 400 & -9.954  & -1.172 & -0.358 & -0.379 & 0.315 & -0.100 \\
16 & 7  & 5  & 0.1  & 450 & -10.103 & -1.197 & -0.325 & -0.310 & 0.266 & -0.084 \\
17 & 7  & 5  & 0.1  & 500 & -10.243 & -1.227 & -0.301 & -0.253 & 0.222 & -0.070 \\
18 & 7  & 5  & 0.1  & 550 & -10.376 & -1.258 & -0.278 & -0.202 & 0.178 & -0.057 \\
19 & 7  & 5  & 0.1  & 600 & -10.495 & -1.280 & -0.266 & -0.170 & 0.152 & -0.049 \\
20 & 7  & 5  & 0.1  & 650 & -10.610 & -1.296 & -0.260 & -0.141 & 0.135 & -0.044 \\
21 & 7  & 5  & 0.1  & 700 & -10.712 & -1.315 & -0.249 & -0.110 & 0.111 & -0.038 \\
22 & 9  & 0  & 0.1  & 400 & -9.949  & -1.107 & -0.632 & -0.720 & 0.733 & -0.208 \\
23 & 9  & 0  & 0.1  & 450 & -10.092 & -1.143 & -0.564 & -0.673 & 0.629 & -0.166 \\
24 & 9  & 0  & 0.1  & 500 & -10.232 & -1.201 & -0.495 & -0.609 & 0.511 & -0.123 \\
25 & 9  & 0  & 0.1  & 550 & -10.364 & -1.268 & -0.425 & -0.532 & 0.381 & -0.078 \\
26 & 9  & 0  & 0.1  & 600 & -10.488 & -1.304 & -0.387 & -0.506 & 0.316 & -0.053 \\
27 & 9  & 0  & 0.1  & 650 & -10.606 & -1.345 & -0.358 & -0.466 & 0.253 & -0.031 \\
28 & 9  & 0  & 0.1  & 700 & -10.712 & -1.408 & -0.311 & -0.387 & 0.150 & 0.000  \\
29 & 9  & 1  & 0.1  & 400 & -9.990  & -1.109 & -0.632 & -0.718 & 0.731 & -0.207 \\
30 & 9  & 1  & 0.1  & 450 & -10.134 & -1.142 & -0.565 & -0.674 & 0.632 & -0.167 \\
31 & 9  & 1  & 0.1  & 500 & -10.273 & -1.202 & -0.497 & -0.606 & 0.511 & -0.124 \\
32 & 9  & 1  & 0.1  & 550 & -10.406 & -1.269 & -0.422 & -0.531 & 0.379 & -0.078 \\
33 & 9  & 1  & 0.1  & 600 & -10.530 & -1.306 & -0.382 & -0.503 & 0.311 & -0.051 \\
34 & 9  & 1  & 0.1  & 650 & -10.648 & -1.343 & -0.361 & -0.468 & 0.258 & -0.032 \\
35 & 9  & 1  & 0.1  & 700 & -10.753 & -1.408 & -0.313 & -0.387 & 0.153 & -0.001 \\
36 & 9  & 5  & 0.05 & 400 & -10.093 & -1.216 & -0.416 & -0.334 & 0.370 & -0.126 \\
37 & 9  & 5  & 0.05 & 450 & -10.253 & -1.211 & -0.407 & -0.316 & 0.358 & -0.117 \\
38 & 9  & 5  & 0.05 & 500 & -10.404 & -1.226 & -0.394 & -0.289 & 0.328 & -0.104 \\
39 & 9  & 5  & 0.05 & 550 & -10.548 & -1.243 & -0.373 & -0.267 & 0.291 & -0.088 \\
40 & 9  & 5  & 0.05 & 600 & -10.678 & -1.260 & -0.360 & -0.247 & 0.263 & -0.077 \\
41 & 9  & 5  & 0.05 & 650 & -10.804 & -1.277 & -0.347 & -0.227 & 0.237 & -0.068 \\
42 & 9  & 5  & 0.05 & 700 & -10.918 & -1.298 & -0.327 & -0.203 & 0.200 & -0.055 \\
43 & 9  & 5  & 0.1  & 400 & -10.073 & -1.191 & -0.435 & -0.448 & 0.445 & -0.139 \\
44 & 9  & 5  & 0.1  & 450 & -10.218 & -1.205 & -0.423 & -0.403 & 0.421 & -0.131 \\
45 & 9  & 5  & 0.1  & 500 & -10.359 & -1.242 & -0.400 & -0.344 & 0.368 & -0.114 \\
46 & 9  & 5  & 0.1  & 550 & -10.490 & -1.278 & -0.379 & -0.294 & 0.317 & -0.098 \\
47 & 9  & 5  & 0.1  & 600 & -10.611 & -1.306 & -0.363 & -0.260 & 0.281 & -0.086 \\
48 & 9  & 5  & 0.1  & 650 & -10.727 & -1.331 & -0.355 & -0.227 & 0.252 & -0.078 \\
49 & 9  & 5  & 0.1  & 700 & -10.831 & -1.356 & -0.338 & -0.197 & 0.216 & -0.066 \\
50 & 9  & 5  & 0.2  & 400 & -10.049 & -1.147 & -0.494 & -0.650 & 0.599 & -0.171 \\
51 & 9  & 5  & 0.2  & 450 & -10.173 & -1.200 & -0.454 & -0.544 & 0.518 & -0.150 \\
52 & 9  & 5  & 0.2  & 500 & -10.293 & -1.262 & -0.421 & -0.446 & 0.436 & -0.129 \\
53 & 9  & 5  & 0.2  & 550 & -10.408 & -1.314 & -0.398 & -0.368 & 0.370 & -0.112 \\
54 & 9  & 5  & 0.2  & 600 & -10.515 & -1.360 & -0.375 & -0.303 & 0.312 & -0.096 \\
55 & 9  & 5  & 0.2  & 650 & -10.618 & -1.390 & -0.372 & -0.257 & 0.282 & -0.090 \\
56 & 9  & 5  & 0.2  & 700 & -10.711 & -1.431 & -0.345 & -0.199 & 0.220 & -0.072 \\
57 & 9  & 5  & 0.4  & 400 & -10.039 & -1.100 & -0.639 & -0.963 & 0.883 & -0.235 \\
58 & 9  & 5  & 0.4  & 450 & -10.127 & -1.178 & -0.570 & -0.799 & 0.740 & -0.200 \\
59 & 9  & 5  & 0.4  & 500 & -10.221 & -1.267 & -0.511 & -0.643 & 0.602 & -0.166 \\
60 & 9  & 5  & 0.4  & 550 & -10.316 & -1.349 & -0.463 & -0.506 & 0.479 & -0.137 \\
61 & 9  & 5  & 0.4  & 600 & -10.404 & -1.400 & -0.444 & -0.419 & 0.413 & -0.122 \\
62 & 9  & 5  & 0.4  & 650 & -10.490 & -1.445 & -0.428 & -0.340 & 0.352 & -0.108 \\
63 & 9  & 5  & 0.4  & 700 & -10.567 & -1.491 & -0.398 & -0.263 & 0.276 & -0.089 \\
64 & 9  & 5  & 0.8  & 400 & -10.084 & -1.126 & -0.972 & -1.344 & 1.364 & -0.361 \\
65 & 9  & 5  & 0.8  & 450 & -10.129 & -1.192 & -0.844 & -1.150 & 1.143 & -0.302 \\
66 & 9  & 5  & 0.8  & 500 & -10.185 & -1.283 & -0.740 & -0.952 & 0.933 & -0.248 \\
67 & 9  & 5  & 0.8  & 550 & -10.251 & -1.369 & -0.660 & -0.778 & 0.756 & -0.204 \\
68 & 9  & 5  & 0.8  & 600 & -10.312 & -1.436 & -0.607 & -0.640 & 0.625 & -0.172 \\
69 & 9  & 5  & 0.8  & 650 & -10.379 & -1.489 & -0.569 & -0.526 & 0.523 & -0.148 \\
70 & 9  & 5  & 0.8  & 700 & -10.438 & -1.541 & -0.526 & -0.419 & 0.416 & -0.122 \\
71 & 9  & 15 & 0.1  & 400 & -9.963  & -1.175 & -0.386 & -0.412 & 0.376 & -0.119 \\
72 & 9  & 15 & 0.1  & 450 & -10.112 & -1.194 & -0.367 & -0.353 & 0.341 & -0.108 \\
73 & 9  & 15 & 0.1  & 500 & -10.254 & -1.227 & -0.345 & -0.298 & 0.295 & -0.094 \\
74 & 9  & 15 & 0.1  & 550 & -10.385 & -1.256 & -0.327 & -0.255 & 0.254 & -0.081 \\
75 & 9  & 15 & 0.1  & 600 & -10.507 & -1.279 & -0.317 & -0.226 & 0.229 & -0.072 \\
76 & 9  & 15 & 0.1  & 650 & -10.621 & -1.305 & -0.305 & -0.189 & 0.194 & -0.062 \\
77 & 9  & 15 & 0.1  & 700 & -10.724 & -1.329 & -0.290 & -0.158 & 0.161 & -0.052 \\
78 & 9  & 50 & 0.1  & 400 & -9.462  & -1.139 & -0.382 & -0.475 & 0.420 & -0.131 \\
79 & 9  & 50 & 0.1  & 450 & -9.612  & -1.169 & -0.358 & -0.402 & 0.365 & -0.115 \\
80 & 9  & 50 & 0.1  & 500 & -9.754  & -1.209 & -0.335 & -0.336 & 0.307 & -0.099 \\
81 & 9  & 50 & 0.1  & 550 & -9.888  & -1.252 & -0.310 & -0.274 & 0.245 & -0.081 \\
82 & 9  & 50 & 0.1  & 600 & -10.010 & -1.286 & -0.295 & -0.229 & 0.199 & -0.067 \\
83 & 9  & 50 & 0.1  & 650 & -10.122 & -1.311 & -0.288 & -0.197 & 0.169 & -0.058 \\
84 & 9  & 50 & 0.1  & 700 & -10.223 & -1.337 & -0.280 & -0.168 & 0.138 & -0.048 \\
85 & 14 & 5  & 0.1  & 400 & -10.281 & -1.243 & -0.598 & -0.451 & 0.601 & -0.196 \\
86 & 14 & 5  & 0.1  & 450 & -10.422 & -1.266 & -0.585 & -0.399 & 0.569 & -0.183 \\
87 & 14 & 5  & 0.1  & 500 & -10.559 & -1.301 & -0.570 & -0.349 & 0.531 & -0.170 \\
88 & 14 & 5  & 0.1  & 550 & -10.689 & -1.337 & -0.553 & -0.300 & 0.487 & -0.156 \\
89 & 14 & 5  & 0.1  & 600 & -10.810 & -1.365 & -0.538 & -0.267 & 0.457 & -0.146 \\
90 & 14 & 5  & 0.1  & 650 & -10.927 & -1.396 & -0.521 & -0.224 & 0.419 & -0.135 \\
91 & 14 & 5  & 0.1  & 700 & -11.036 & -1.429 & -0.487 & -0.182 & 0.365 & -0.119 \\
\hline
\label{tab:fitparamwinter}
\end{longtable}


\bsp 
\label{lastpage}
\end{document}